\documentclass[prd,nofootinbib,twocolumn,showpacs]{revtex4}
\usepackage{graphics}
\usepackage{bm}

\def\be{\begin{equation}}
\def\ee{\end{equation}}
\def\ba{\begin{eqnarray}}
\def\ea{\end{eqnarray}}

\begin{document}

\title{\large \bf Observational constraints on cosmic strings: Bayesian analysis
in a three dimensional parameter space}
\author{Levon Pogosian$^1$, Mark Wyman$^{2,3}$, and Ira Wasserman$^{2,3}$ \vspace{0.2cm}}
\affiliation{$^1$ Institute of Cosmology, Department of Physics and Astronomy,
Tufts University, Medford, MA 02155, USA \\
$^2$ Laboratory for Elementary Particle Physics, Cornell University,
Ithaca, NY 14853, USA \\
$^3$ Center for Radiophysics and Space Research, Cornell University,
Ithaca, NY 14853, USA}
\date{today}


\begin{abstract}

Current data exclude cosmic strings as the primary source of
primordial density fluctuations. However, in a wide class of
inflationary models, strings can form at later stages of inflation
and have potentially detectable observational signatures. We study
the constraints from WMAP and SDSS data on the fraction of
primordial fluctuations sourced by local cosmic strings. The
Bayesian analysis presented in this brief report is restricted to
the minimal number of parameters. Yet it is useful for two
reasons. It confirms the results of \cite{PTWW03} using an
alternative statistical method. Secondly, it justifies the more
costly multi-parameter analysis. Already, varying only three
parameters -- the spectral index and the amplitudes of the adiabatic
and string contributions -- we find that the upper bound on the
cosmic string contribution is of order $10\%$. We expect that
the full multi-parameter study, currently underway \cite{PWW04},
will likely loosen this bound.
\end{abstract}


\

\maketitle

The idea that inflationary cosmology might lead to cosmic string
formation is not new \cite{kofman} and has received new impetus
from brane inflation scenarios suggested by superstring theory
\cite{dvali-tye,jst,costring,DV02,jst2,DV03,Polchinski03} . In
these models, inflation can arise during the collisions of branes
that coalesce to form, ultimately, the brane on which we live
\cite{dvali-tye,burgess,rabadan,collection}. It is possible to
build brane inflation models that predict adiabatic temperature
and dark matter fluctuations capable of reproducing all currently
available observations. A seemingly unavoidable outcome of brane
inflation, however, is the production of a network of local cosmic
strings \cite{jst,costring}, whose effects on cosmological
observables range from negligible to substantial, depending on
the specific brane inflationary scenario \cite{jst2,DV03}. As the
precision of cosmological observations increases, we might hope to
be able to distinguish among numerous presently viable models of
inflation by the properties of the cosmic strings they predict.

The fluctuations resulting from brane inflation are expected to be an
incoherent superposition of contributions from adiabatic perturbations
initiated by curvature fluctuations and active perturbations
induced by the decaying cosmic string network. The resulting
CMB temperature power spectrum can be written as
\be
C_l=WC_l^{\rm adiabatic}+BC_l^{\rm strings},
\label{clsuperpos}
\ee
where $W$ and $B$ are weighting factors. Analogous expressions hold for
matter density and polarization power spectra. In Eq. (\ref{clsuperpos}),
the weight factors $W$ and $B$ determine the relative importance of the
adiabatic and cosmic string contributions. We choose the weight factor
$W$ so that $W=1$ for the WMAP's best fit power law model \cite{wmap_spergel}
with no cosmic strings.

The scaling properties of the cosmic string networks in brane
inflationary scenarios are different from those of the more
familiar 3+1 D field theoretical strings, since intercommutation
probabilities are smaller as a consequence of the existence of
extra dimensions \cite{jst2,DV03}. As argued in \cite{PTWW03},
observational constraints on the amplitude of string-induced CMB
perturbations place limits on $G\mu/\sqrt{\lambda}$, where $\mu$
is the string tension and $\lambda \le 1$ is a dimensionless
measure of the intercommutation rate. The constraint reported in
\cite{PTWW03}, based on WMAP
\cite{wmap_bennett,wmap_spergel,wmap_verde} and 2dF data
\cite{2dF}, was $G\mu\lesssim 1.3\times
10^{-6}\sqrt{B\lambda/0.1}$, where $B$, defined by
Eq.~(\ref{clsuperpos}), measures the importance of perturbations
induced by cosmic strings. It was argued that, conservatively, the
currently available data still permit $B\lesssim 0.1$. This bound
on the parameter $B$ was found by minimizing the $\chi^2$ with
respect to variations in the parameters $B$, $W$, and $n_s$, the
spectral index of the adiabatic fluctuations power spectrum. The
uncertainty in the minimum $\chi^2$ was translated into an upper
bound on $B$. This statistical procedure was not rigorous,
although the derived bound should be reasonably secure.

In this paper we re-calculate the bound on $B$, varying the same
parameters as in \cite{PTWW03}, but using a rigorous Bayesian
likelihood analysis. Except for the statistical methods used, and
for fitting to the SDSS instead of the 2dFGRS power spectrum, our
setup is identical to the one used in \cite{PTWW03} and the
interested reader can find most of the details there. Here we only
mention that $C_l^{\rm strings}$ were computed using a modified
version \cite{levon} of CMBFAST \cite{cmbfast} which uses as its
active source the energy-momentum tensor components calculated
using the local string network model originally suggested in
\cite{aletal} and further improved in \cite{levon,gangui}.
$C_l^{\rm adiabatic}$ were computed using a recent of version of
CMBFAST. As in \cite{PTWW03}, we vary the value of the spectral
index $n_s$, while holding the following parameters fixed at their
best-fit values as determined by WMAP \cite{wmap_spergel}:
$\Omega_m h^2 = 0.14$, $\Omega_b h^2 = 0.024$, $\Omega_{\Lambda} =
1 - \Omega_m - \Omega_b$, $h = 0.72$, $\tau= 0.166$. Our rationale
is that cosmic strings are only a small correction to the WMAP
best-fit. Elsewhere \cite{PWW04}, we shall present a more
complete study in which we also allow $\Omega_m h^2,\,\Omega_b
h^2,\, \tau$ and $h$ to vary as well.

According to Bayes's theorem, the posterior distribution for $W$,
$B$ and $n_s$ is given by \be P(W,B, n_s\vert D,M)={P(W,B,
n_s\vert M){\cal L}(D\vert W,B, n_s, M)\over P(D\vert M)}~, \ee
where ${\cal L}(D\vert W,B, n_s, M)$ is the likelihood, $M$
denotes our model and $D$ stands for data. We use WMAP's CMB
temperature anisotropy power spectrum ($TT$) and the
temperature-polarization cross-correlation ($TE$) together with
the power spectrum ($P(k)$) from the SDSS experiment. For $TT$ and
$TE$ we have used the likelihood function provided by the WMAP
team \cite{wmapwebsite, wmap_verde} and for $P(k)$ the SDSS
likelihood code provided by Max Tegmark \cite{tegmark,maxwebsite}.
The combined $TT+TE+P(k)$ likelihood is simply the product of
individual likelihoods.

\begin{figure}[tbp]
\scalebox{0.3}{\includegraphics{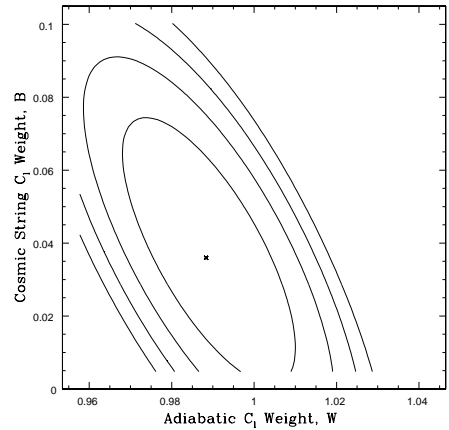}}
\caption{\label{fig:wbcont} $\Delta(\ln({\cal L}))$ = 1, 2, 3, and 4 contours
after the likelihood grid has been
marginalized over $n_s$. The maximum likelihood model is marked.}
\end{figure}

\begin{figure}[tbp]
\scalebox{0.3}{\includegraphics{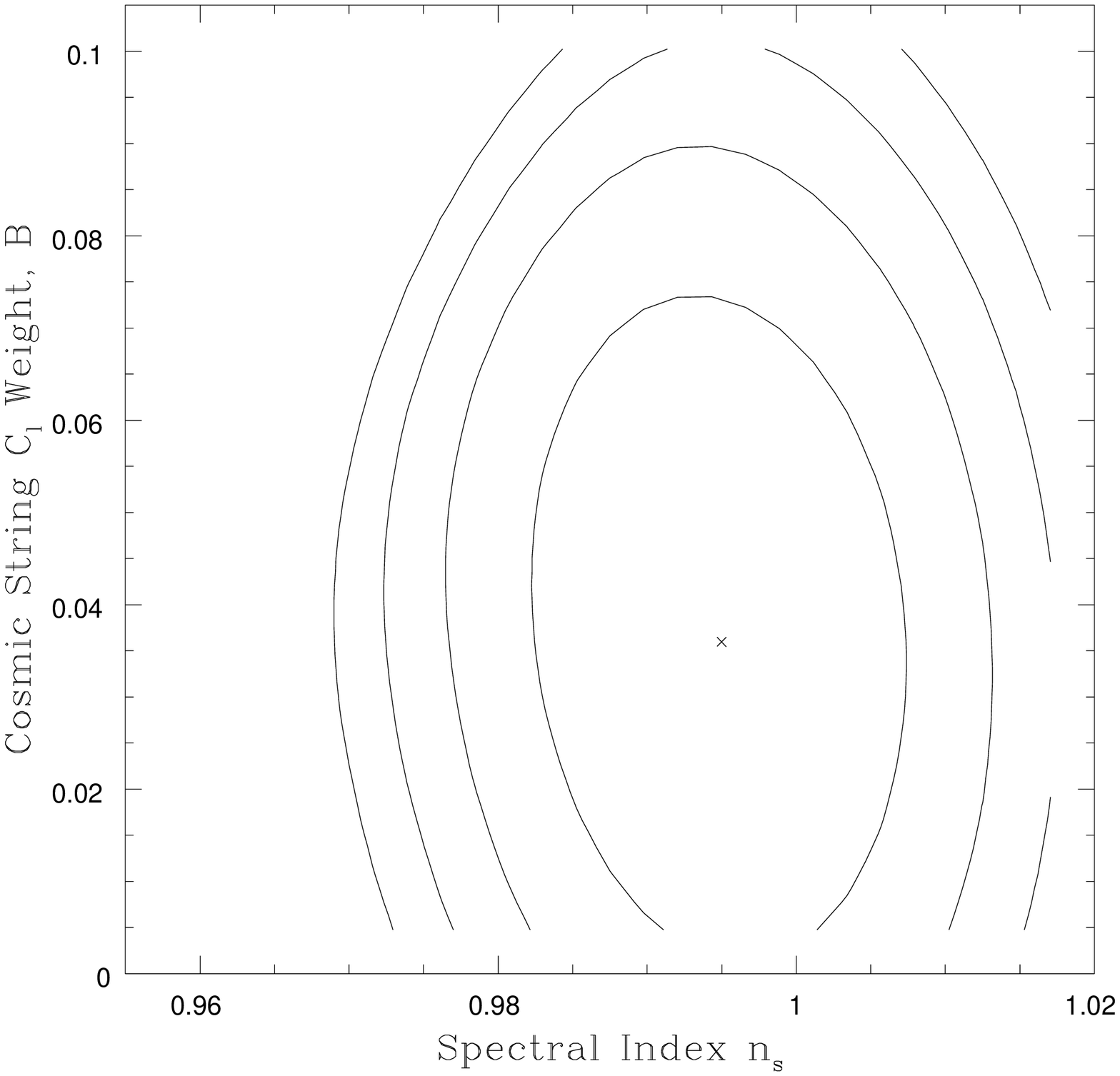}}
\caption{\label{fig:nbcont} $\Delta(\ln({\cal L}))$  = 1, 2, 3, and 4 contours
after the likelihood grid has been marginalized over $W$. The maximum likelihood model is marked.}
\end{figure}

\begin{figure}[tbp]
\scalebox{0.3}{\includegraphics{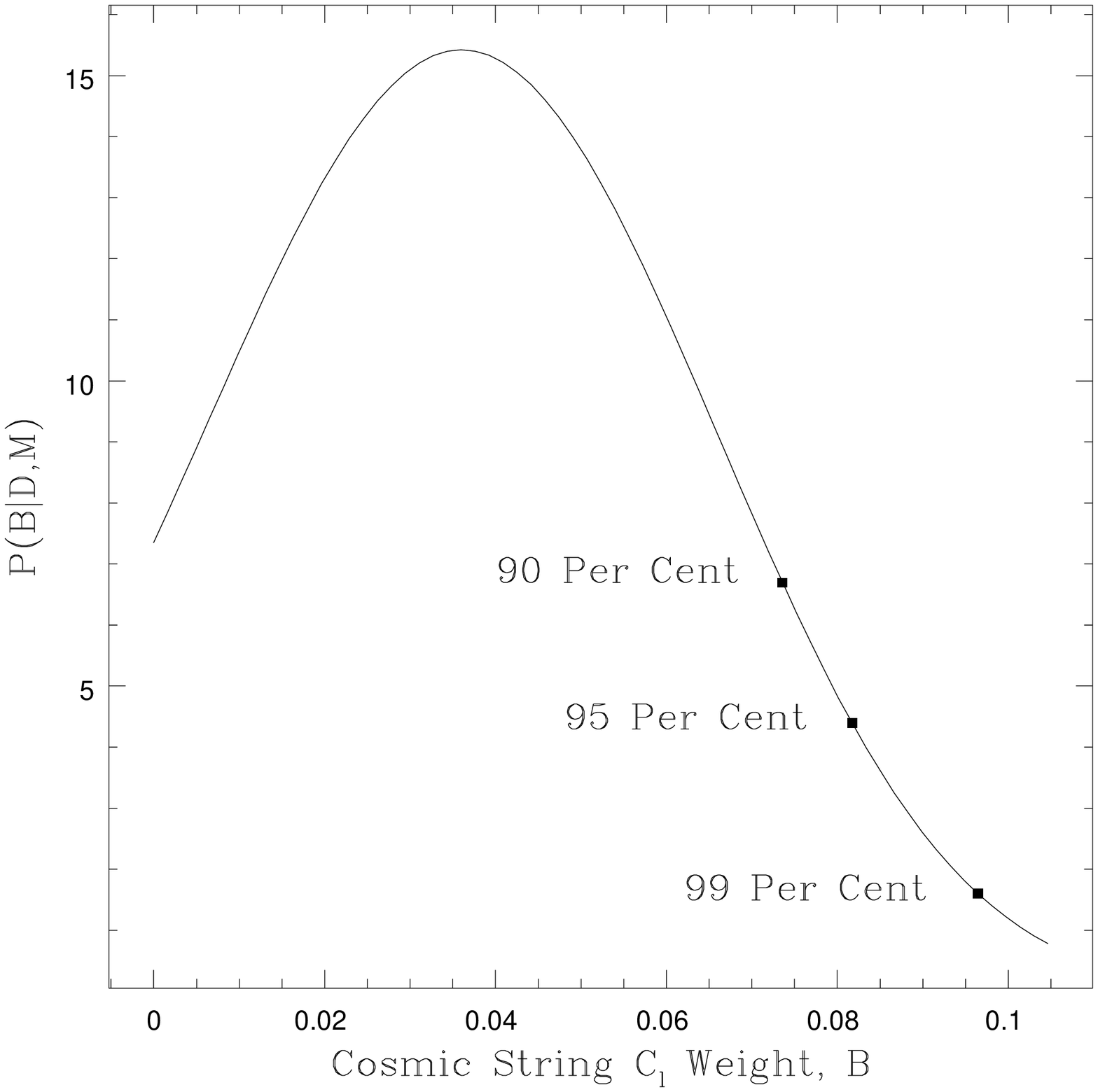}}
\caption{\label{fig:bmarg} This plot shows $P(B|D,M)$
after marginalization over both $W$ and $n_s$. The three dots indicate
the regions containing 90, 95, and 99 per cent of the probability, respectively}
\end{figure}

\begin{figure}[tbp]
\scalebox{0.3}{\includegraphics{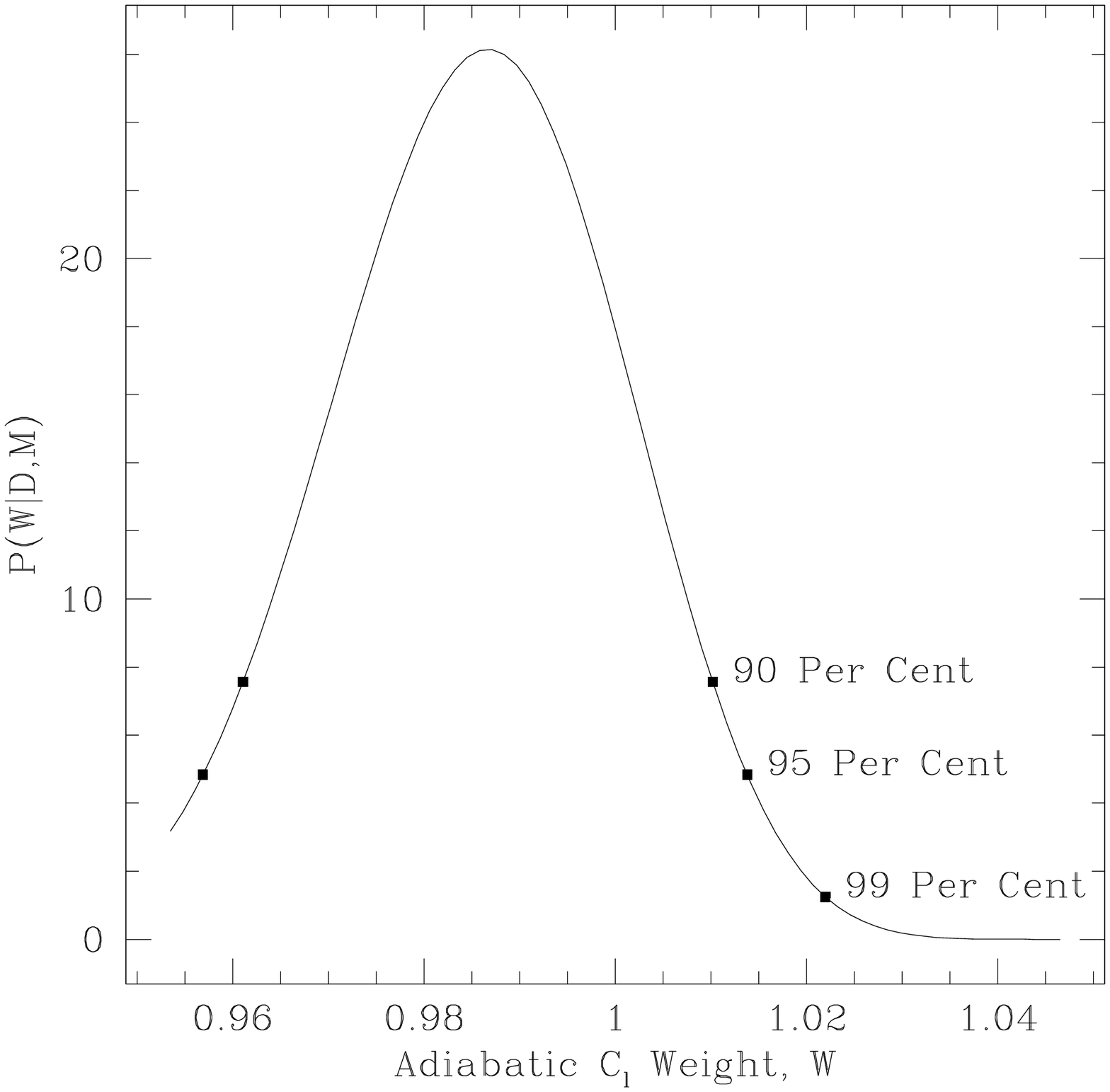}}
\caption{\label{fig:wmarg} This plot shows $P(W|D,M)$
 after marginalization over both $B$ and $n_s$. The dots indicate
the regions containing 90, 95, and 99 per cent of the probability, respectively}
\end{figure}

\begin{figure}[tbhp]
\scalebox{0.3}{\includegraphics{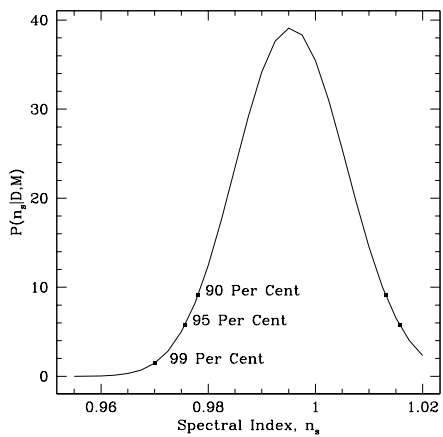}}
\caption{\label{fig:nsmarg} This plot shows $P(n_s|D,M)$
after marginalization over both $W$ and $B$. The dots indicate
the regions containing 90, 95, and 99 per cent of the probability, respectively}
\end{figure}

We choose flat priors, $P(W,B,n_s)$ for parameters $W$, $B$ and $n_s$ with ranges
$W \in [0.95, 1.05], B \in [0.0,0.105]$, and $n_s \in [0.955, 1.02]$.
 We then survey this parameter space
using a $73 \times 65 \times 27$ grid for $W, B$ and $n_s$, respectively.
At each grid point we calculate four predicted spectra from our model: $C_l^{TT}$,
$C_l^{TE}$, $C_l^{EE}$, and $P(k)$.
The SDSS power spectrum is determined up to an overall bias factor, so the
bias has to be included into our analysis as a fourth parameter. The bias and
the effect of the spectral distortions on $P(k)$ can
be parametrized via an overall multiplicative factor, which we vary over the range
$[1.50,1.86]$  -- a range large enough to include the maximum likelihood bias value for
every grid point -- and over which we marginalize.  After compiling the
full likelihood grid, we marginalize over $W$
and $n_s$ separately and plot the resulting likelihood contours.
In Figs.~\ref{fig:wbcont} and \ref{fig:nbcont} we show the contours in the
$(W,B)$ and $(n_s,B)$ planes respectively.
Finally, we marginalize over both $W$ and $n_s$ to find the posterior for
$B$, shown in Fig.~\ref{fig:bmarg}. In a similar manner, we have also
generated the posteriors for $W$ (Fig.~\ref{fig:wmarg}) and
$n_s$ (Fig.~\ref{fig:nsmarg}). The posterior results for $W$ can be interpreted
in light of Fig.~\ref{fig:bmarg}'s results for $B$, especially with the help of
Fig.~\ref{fig:wbcont}. We see that there is something of a degeneracy between $W$ and $B$,
so that an increase in $B$ is compensated by a decrease in $W$. Nevertheless,
our determination, $W = 0.985 \pm 0.025$, has a much smaller range than
that found by WMAP with $B=0$: $W=1.0\pm 0.11$  \cite{wmap_spergel} . For the spectral
index, we find $n_s = 0.995\pm0.0165$. For comparison, Spergel et al.
\cite{wmap_spergel} find $n_s = 0.99 \pm 0.04$ but vary many more parameters.
We note, from Fig.~\ref{fig:nbcont}, that the approximate $\Delta \ln({\cal L})=1$ range
for $n_s$ is $\approx 0.995\pm0.005$ at $B=0$. We therefore suspect
that the relatively narrow range we find for $n_s$ could broaden
 when we vary $\Omega_m,\,\Omega_b,\, \tau$ and $h$ in addition to
 $W,\,B,$ and $n_s$.

Our principal result is Fig.~\ref{fig:bmarg}, the posterior for our cosmic string
weighting parameter, $B$. We note that, despite our use of the unadjusted
WMAP best-fit parameters for all cosmological factors save $n_s$,
our additional cosmic string degree of freedom  still improves the fit to the data,
with the likelihood maximum likelihood falling
at $B \approx 0.04$. However, at the 90\% confidence level, our results
are also roughly consistent with $B = 0$, while the upper bound at 99\% confidence
is $B\lesssim 0.09$, quite close to what was found much less rigorously in \cite{PTWW03}. 
This upper bound might still permit observable signatures such as pulsar timing \cite{timing}.
gravitational wave bursts \cite{bursts}, and distinctive double lensing events \cite{lensing,lens2}.
 That the peak for $B$ is not at zero, however, underscores the necessity of
 varying the other cosmological parameters,
 $\tau,\,h,\, \Omega_m,$ and $\Omega_b$ as well. We have begun work on this multi-parameter
 analysis using the Markov Chain Monte Carlo method \cite{PWW04}.

\acknowledgments We thank Henry Tye for helpful discussions, Licia
Verde for assistance with using the WMAP likelihood code and Max
Tegmark for writing an easy to use SDSS likelihood code and making
it publicly available \cite{maxwebsite}. LP would like to thank
Ken Olum and Alex Vilenkin for useful conversations. This research
is partially supported by NSF Grant No. AST 0307273 (I.W.). M.W is
supported by the NSF Graduate Fellowship.

\end{document}